\documentclass[twoside]{article}
\usepackage{qic,epsfig}
\newcommand{\PRL}{Phys. Rev. Lett.}
\newcommand{\PRA}{Phys. Rev. A}
\newcommand{\JPD}{J. Phys. D: Appl. Phys.}

\begin{document}

\setlength{\textheight}{8.0truein}

\runninghead{Fabrication of micro-magnetic traps for cold neutral atoms}
            {Benjamin Lev}

\normalsize\textlineskip

\thispagestyle{empty}

\setcounter{page}{1}

\vspace*{0.88truein}

\alphfootnote

\fpage{1}

\centerline{\bf FABRICATION OF MICRO-MAGNETIC TRAPS FOR}
\vspace*{0.035truein} \centerline{\bf COLD NEUTRAL ATOMS}
\vspace*{0.37truein} \centerline{\footnotesize BENJAMIN LEV}
\vspace*{0.015truein} \centerline{\footnotesize\it Norman Bridge
Laboratory of Physics 12-33, California Institute of Technology}
\baselineskip=10pt \centerline{\footnotesize\it Pasadena,
California 91125, USA} \centerline{\footnotesize\it
benlev@caltech.edu}\vspace*{0.225truein}

\vspace*{0.21truein}

\abstracts{Many proposals for quantum information processing require precise
control over the motion of neutral atoms, as in the manipulation of coherent
matter waves or the confinement and localization of individual atoms. Patterns
of micron-sized wires, fabricated lithographically on a flat substrate, can
conveniently produce large magnetic-field gradients and curvatures to trap
cold atoms and to facilitate the production of Bose-Einstein condensates. This
paper describes tools and techniques for the construction of such
devices.}{}{}

\vspace*{10pt}

\vspace*{1pt}\textlineskip

\section{Introduction}

Cold samples of neutral atoms and Bose-Einstein condensates have
become readily available using the techniques of laser cooling and
trapping~\cite{Metcalf99}, and it has been widely recognized that
cold atoms are a rich resource for experiments in quantum
information science. For many proposals, however, quantum control
of the atomic motional degrees of freedom is essential. For
example, quantum computation in a cavity QED setting or through
controlled cold collisions requires the ability to trap and
control single atoms in the Lamb-Dicke
regime~\cite{Zoller95,Calarco,Briegel}. In 1995, Weinstein and
Libbrecht noted that micron-sized wires, fabricated on a
substrate, are capable of producing the large magnetic field
gradients and curvatures required for trapping atoms in the
Lamb-Dicke regime~\cite{Libbrecht}. Westervelt {\it et al.}, in
1998, succeeded in fabricating the wire patterns used in the trap
designs of Weinstein and Libbrecht~\cite{Westervelt98a}. These
microwire devices, now commonly known as atom
chips~\cite{Schmiedmayer00b}, have been used to great success in
atom optics and in the production of Bose-Einstein condensates
(BEC), and are promising tools not just for quantum computation,
but for atom interferometry, cavity QED, and the study of cold
collisions as well.  In this paper we describe techniques, which
have been adapted from the standard lore of microfabrication, for
fabricating this increasingly important tool for atomic physics
and quantum optics.

Atom optical elements, such as mirrors, waveguides, splitters,
traps, and conveyor belts have been demonstrated using atom
chips~\cite{Jakob01a,Folman02,Anderson01,Hindsrev,Westervelt98b,Hannaford02,Lev03}.
Cesium cold collisions in the presence of light have been studied
using a magnetic microtrap~\cite{Levpre}, and the use of fiber
gap~\cite{Hinds02,Jakob03} and microsphere cavities~\cite{Jakob03}
for on-chip atom detection is being explored. Ion trap experiments
are now using substrates with microfabricated electric pads for
the purpose of controlling ion
position~\cite{Wineland02a,Wineland02b}.

On-chip production of a BEC has been one of the most successful uses of the
atom chip thus far~\cite{Jakob01b,Zimmermann01,Ketterle03}. Ioffe traps formed
from microwires can produce extremely large trap compressions that enhance the
efficiency of evaporative cooling. Consequently, condensate production time
can be reduced from one minute to approximately ten seconds~\cite{Jakob01b},
and MOT loading can occur from a thermal vapor in a glass cell with a vacuum
of only a few $10^{-10}$ Torr. All of the required magnetic fields can be
produced on-chip~\cite{Schmiedmayer03,Schmiedmayer00}, removing the necessity
of large, high power external coils. The atom chip greatly miniaturizes BEC
production and will enable the integration of matter waves with chip-based
atom optics and photonics.

Another exciting avenue of research involves the use of an atom
chip to trap, in the Lamb-Dicke regime, one or more atoms in the
mode of a high finesse cavity. The combination of magnetic
microtraps and photonic bandgap (PBG) cavities would be an
excellent cavity QED system for the implementation of scalable
quantum computation, or for the study of continuous measurement
and quantum-limited feedback. One technical proposal involves the
integration of a PBG cavity with an Ioffe trap formed from
microwires patterned on the same surface~\cite{Mabuchi01}. The
combination of small mode volume and modest optical quality factor
that should be obtainable with PBG structures would enable strong
atom-cavity coupling. This would be an interesting alternative to
present experiments that utilize a Far Off Resonance Trap (FORT)
to confine atoms inside optical Fabry-Perot
cavities~\cite{McKeever03}. Several PBG cavities, each with an
independent microwire trap, could be fabricated on the same
substrate and coupled together with a network of line-defect
optical waveguides.

Atom chips exploit the interaction potential, $V=-\vec{\mu}\cdot\vec{B}$,
between an atom's magnetic moment, $\vec{\mu}$, and a wire's magnetic field,
$\vec{B}$, to trap or guide weak-field seeking states of a neutral atom.  In
general, the field's magnitude, gradient, and curvature scale as $I/r$,
$I/r^2$, and $I/r^3$, respectively, where $I$ is the wire's current and $r$ is
its characteristic dimension. Microscopic wire patterns maximize field
gradients and curvatures while keeping power dissipation to a minimum.
Experiments involve ultra-high vacuum chambers wherein atoms are trapped and
cooled near the vicinity of the atom chip's confining magnetic potentials.

\section{Fabrication Challenges and Constraints}
\begin{figure} [htbp]
\centerline{\epsfig{file=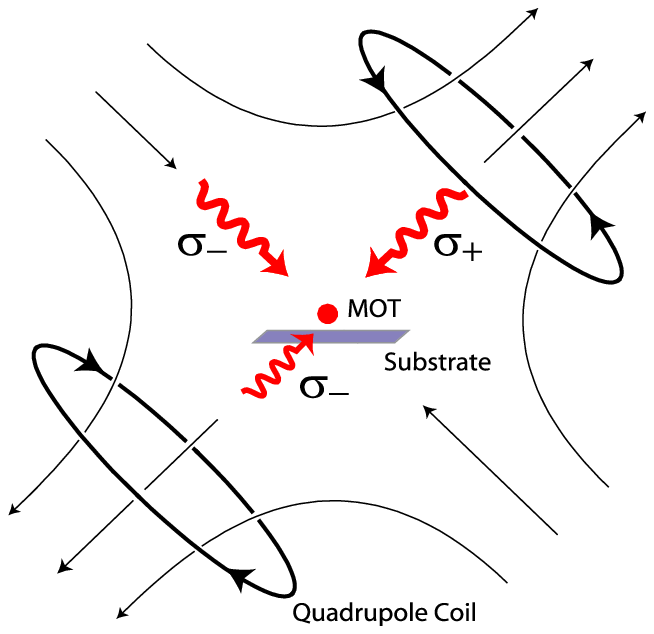, width=6.2cm}}
\vspace*{13pt} \fcaption{\label{fig:mMOT} Diagram of the mirror
MOT experimental set-up. A quadrupole field and two 45$^\circ$
laser beams and one retroreflected grazing beam form a MOT 1.5 to
4 mm above the atom mirror.}
\end{figure}

\noindent Fabrication of atom chips poses several challenges in
addition to those encountered in standard photolithography.  Many
applications require the wires to be a couple microns wide by a
few microns tall and spaced only a few microns from one another.
One micron resolution is near the limit of standard
photolithography, and much care must be taken to accurately
produce these micron-sized wires.  Wires with widths much less
than a micron are of limited usefulness since they become limited
to the same maximum current density as micron-sized
wires~\cite{Jakob02}. Further fabrication complications arise from
the need to trap the atoms near the substrate's surface, and the
need to connect the microwires to macroscopic leads without
blocking optical access. A common technique for trapping atoms
near the substrate surface, the mirror magneto-optical trap
(MMOT), requires that this surface be an optical mirror as well as
the support surface for the microwires (see
figure~\ref{fig:mMOT})~\cite{Jakob99}.  The substrate surface
needs to be larger than 5 to 10 cm$^2$ to accommodate the
reflected trapping beams as well as to allow the pads for
macroscopic wire contacts to be outside of the mirror area and not
blocking the optical access needed for the trapping, imaging, and
pumping beams. Consequently, the wire pattern must be flawless
over an exceptionally large surface area: during fabrication one
must be extremely careful that no dust or surface defects break or
short the wires.

The major fabrication challenge lies in increasing the height of
the wires to a few microns. Even the smallest wires need to
support up to an amp of current, and consequently, the
cross-sectional area of the wire must be maximized.  This reduces
wire resistance and limits the heating that causes wire breakdown.
Moreover, attention must be paid to the thermal conductivity of
the substrate and mounting system to ensure sufficient power
dissipation. Sapphire or polished aluminum nitride (AlN)
substrates provide sufficient thermal conductivity, but are
slightly trickier to use for fabrication than more standard
substrates.

\begin{figure} [htbp]
\centerline{\epsfig{file=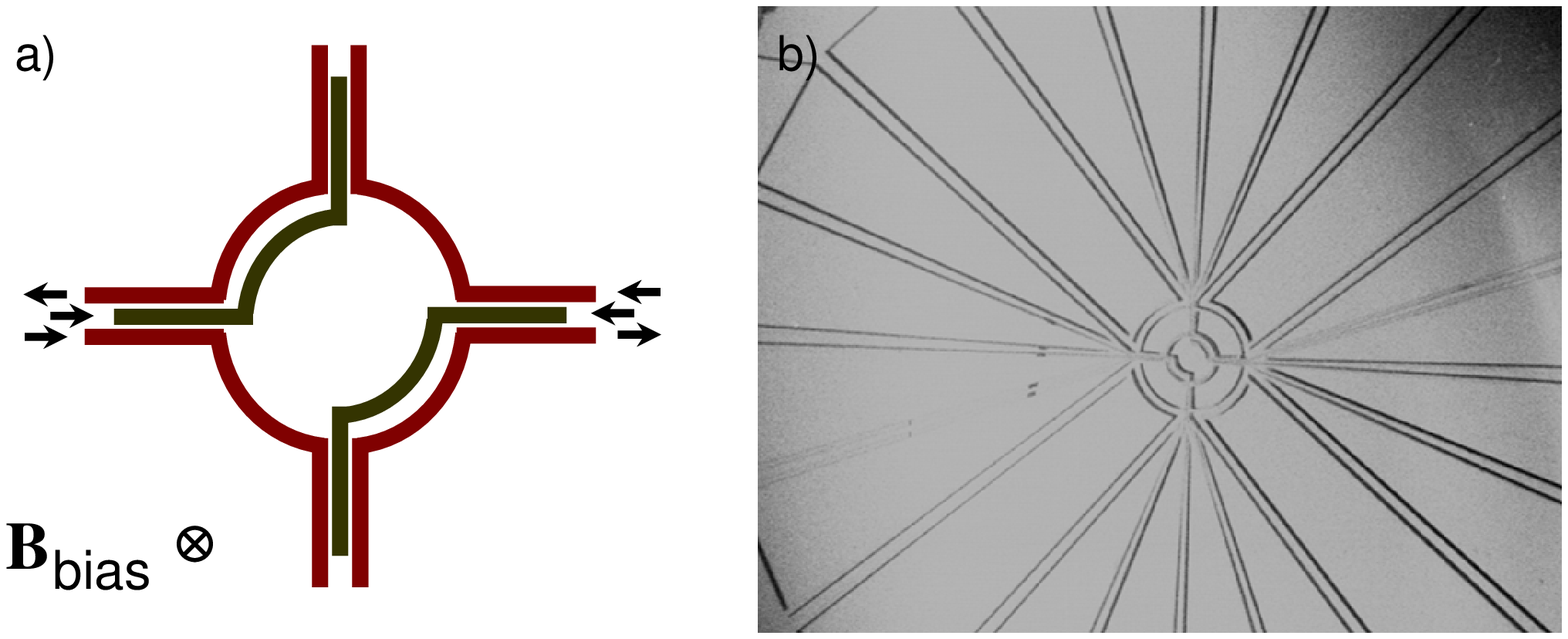, width=11.2cm}} \vspace*{13pt}
\fcaption{\label{fig:Ioffe} The planar Weinstein and
Libbrecht-style Ioffe trap. a) When combined with an opposing bias
field, this wire pattern produces a 3D harmonic potential above
the substrate with a non-zero field at the trap
center~\cite{Libbrecht}.  b) A planar Ioffe trap with an on-chip
bias coil fabricated with gold on sapphire using the lift-off
method.  In the sample shown here, the wire height is 1.5 $\mu$m
and the minimum wire width is 10 $\mu$m. The gold between the
wires forms a mirror for creating a mirror MOT.}
\end{figure}

The use of microwires to create an Ioffe trap illustrates these
challenges. The wire pattern shown in figures~\ref{fig:Ioffe}(a)
and (b) creates a 3D harmonic trap when combined with a
perpendicular homogenous bias field~\cite{Libbrecht}. Unlike a
quadrupole trap, the Ioffe trap has a non-zero field at the trap
center and thus does not suffer from Majorana spin-flip losses. An
atom is confined within the Lamb-Dicke regime when its recoil
energy is less then the trap's vibrational level spacing
($\eta=(E_{recoil}/E_{vib})^{1/2}<1$), and for a cesium atom this
occurs when the trap curvature exceeds $2\times10^6$ G/cm$^2$. To
achieve this extremely large field curvature in all three
dimensions, the radius of the wire pattern in
figure~\ref{fig:Ioffe}(a) must be smaller than $\sim30\ \mu$m. For
a trap of inner radius 10 $\mu$m, outer radius 15 $\mu$m, and wire
current $I=1$ A, the curvature and Lamb-Dicke parameter, $\eta$,
at the center of the trap in the axis perpendicular (plane
parallel) to the substrate is $2\times10^8$ G/cm$^2$
($2\times10^{10}$ G/cm$^2$) and $\eta=0.38$ ($\eta=0.11$).  The
closely spaced wires can only be a few microns wide, and even if
fabricated to a height of 2 to 4 microns, the wires would need to
support the large current density of $\sim10^{11}$ A/m$^2$.  The
accommodation of laser beams for atom cooling, loading, and
imaging constrains and complicates the atom chip's design.  The
trap minimum is only 7 $\mu$m from the substrate's surface, and
the mirror patterned on the surface for use with the MMOT must
neither short the Ioffe wires nor extend more than $\sim5\ \mu$m
from the surface. The following sections describe the necessary
fabrication tools and the techniques we use to overcome these
challenges.

\section{The Elements of Atom Chip Fabrication}
\noindent
Microfabrication is a labor intensive process, often
involving several weeks of trial and error to perfect the
fabrication recipe. However, once the process works, five to ten
atom chips can be produced over a span of two to three days.  The
intent of this paper is to provide the researcher who has access
to a standard clean-room enough information to design and
fabricate an atom chip.  We will describe the use of fabrication
instruments and techniques only insofar as they are relevant to
atom chips. Fabrication is not an exact science, and the
techniques described here may not be optimal, but nevertheless
have proven successful for the chips we have fabricated.

In photolithography, UV light shone through a photomask casts
shadows onto photoresist, a light sensitive polymer, which is
coated on the surface of the substrate. Either positive or
negative photoresist may be used, with the primary difference
being that exposed areas of positive photoresist are removed after
developing whereas exposed areas remain in a process using
negative photoresist.  The various fabrication techniques differ
in how the wire metal and photoresist are used to create the wire
patterns. For instance, the wire metal may be either thermally
evaporated into the trenches created in the photoresist, or grown
upward trough the trenches by electroplating onto a seed metallic
layer underneath the photoresist.  The photoresist and unwanted
metal are removed leaving only the desired wire pattern.
Generally, chip fabrication consists of six steps: creating a
photomask containing the desired wire pattern, using
photolithography to transfer the wire pattern to photoresist on a
substrate, thermally evaporating wire material, increasing the
wire height, preparing the surface mirror, and making contacts to
macroscopic wires.  The details and exact order of these steps
vary depending on the specific requirements of the microwire
pattern to be fabricated. For instance, wires wider than 30 $\mu$m
or less than one micron in height may be fabricated with a much
simpler technique than thinner or taller wires.  This section
discusses the steps common to all techniques.  Procedures required
to increase the wire's thickness pertain to individual fabrication
techniques and will be discussed in the next section.

\subsection{The photomask}
\noindent
The photomask is typically a 10 cm square piece of glass
or transparent plastic on which is printed a positive or negative
1:1 image of the wire pattern.  Wire patterns with widths or
spacings less than $\sim30\ \mu$m require a professionally made
chrome mask: one in which the pattern is written with chromium on
a glass plate. We have used the company Photronics, Inc.
(telephone 619-992-8467) to make photomasks from AutoCAD drawings.
Much care must be taken in producing the AutoCAD files since not
all functions are properly converted to the company's file format.
These masks are quite expensive, costing between \$600 and \$800,
but have sub-micron resolution and are typically shipped within a
week. It is possible to purchase a laser writer to produce
in-house photomasks with resolution down to 0.8 $\mu$m. This can
be a cost effective alternative to purchasing individual masks
from companies.

Many commercial printing shops are capable of printing
transparencies with high enough resolution to serve as photomasks
for wire patterns with features larger than $\sim30\ \mu$m. The
line edges are granular on a scale of a few microns, and the UV
exposure time must be adjusted to account for the ink not being
perfectly opaque. However, the one day turn-around, low cost of
$\sim\$20$, and ease of file preparation---only an .eps file is
typically needed---make the transparency photomask quite an
attractive alternative for large features.

\subsection{The substrate}
\noindent As mentioned earlier, the substrate material for the
atom chip should be carefully chosen:  it must be electrically
insulating, highly polished, insusceptible to fractures upon
localized heating, and an excellent thermal conductor. We have
found that both sapphire and AlN substrates satisfy these
requirements. Sapphire substrates 0.5 mm to 2 mm thick with
surface areas of several cm$^2$ may be purchased from companies
such as Meller Optics, Inc. (telephone 800-821-0180) for \$30 to
\$40 apiece.  A surface quality of 80-50 scratch-dig is sufficient
for fabrication.  The thermal conductivity of AlN, $\sim170-180\
\mbox{W}\mbox{m}^{-1}\mbox{K}^{-1}$ at $20^\circ$C, is $\sim4.5$
times higher than that of sapphire~\cite{Jakob02}.  We measured
that the max current density supported by microwires on AlN,
$\sim2\times10^{11}$ A/m$^2$, is a factor of two greater than for
microwires patterned on sapphire. This was measured using
electroplated gold wires of varying cross-sections patterned
exactly the same way on both AlN and sapphire substrates.
Specifically, we used several 3 $\mu$m and 20 $\mu$m wide wires
whose heights ranged from one to three microns. The substrates
were glued to room temperature copper blocks using EPO-TEK H77
(Epoxy Technology, telephone 978-667-3805), a thermally conductive
epoxy.

Sapphire substrates are easier to use for fabrication because
their transparency allows one to detect and avoid defects and dust
during the photolithography process.  Polished AlN substrates may
be purchased in bulk for less than $\sim$ \$75, and unlike
sapphire, AlN substrates can be cleaved with a diamond scorer to
any shape desired.  The polished AlN still has a considerable
amount of surface roughness---one micron wide plateaus a few
hundred nanometers tall are typical---but we found that it is
nevertheless possible to fabricate on this surface perfect three
micron wide wires spaced less than three microns from one another.
The surface bumps simply map directly onto the upper surface of
the wires.

\subsection{Substrate cleaning}
\noindent
Before the photolithography process may begin, the
surface of the substrate must be cleaned to remove all organic
material and dust. Although some of the following steps may seem
unnecessary and ``overkill," investing the time to thoroughly
clean minimizes the chance that after many hours of work, one
discovers that a piece of dirt has broken or shorted a wire.  The
first step is to immerse the substrate in a beaker of ``piranha
etch," sulfuric acid and hydrogen peroxide in a 10:1 volume ratio
brought to 100$^\circ$C on a hot plate for $\sim5$ min. Teflon
coated, flat tipped tweezers are ideal for manipulating
substrates.  After the etch, the substrate should be placed in a
beaker of acetone, heated again to 100$^\circ$C for a few minutes,
and finally inserted into an ultrasound cleaner for few more
minutes. In extreme cases of substrate grime, a cotton tipped
dowel can be used to manually wipe away the dirt.  Acetone leaves
a thin film---and sometimes even particulate---when allowed to dry
on a substrate's surface. It is imperative that one spray
isopropanol (IPA) onto the substrate as it is removed from the
acetone bath. This rinses the surface of acetone and wets it with
IPA which does not quickly dry. The substrate must then be rinsed
with methanol, which is relatively clean and does not leave a
film, and quickly blown dry with an air or nitrogen gun.  It is
crucial that the air jet is aimed almost parallel to the surface
so that the methanol is blown-off rather than dried on the
substrate.  When done correctly, the only remaining dirt particles
will be along the edge of the substrate that is downwind of the
air jet, and not in the center fabrication region.  If the
substrate is reasonably clean after the piranha etch, then the
acetone step (which may actually add some dirt particulate) may be
skipped, and the substrate should instead be immersed in IPA and
placed inside an ultrasound cleaner.

\subsection{Thermal evaporation}
\noindent
Certain fabrication techniques, to be discussed below,
require that a 100 nm metal layer be thermally evaporated before
coating the surface with photoresist.  We take this opportunity to
discuss the thermal evaporation process.  We use gold for the
wires because of its high electrical conductivity, resistance to
corrosion, and ease of evaporation, electroplating, and wet
etching.  To successfully deposit gold on a substrate's surface,
one must first evaporate a 50~\AA~metallic layer that promotes
adhesion between the gold and the sapphire or AlN.  We typically
use chromium, but titanium may also be used. The magnetic effects
from the thin layer of chromium are negligible. In a thermal
evaporator, the substrate is mounted in a vacuum chamber facing a
tungsten crucible positioned a few tens of centimeters below. The
crucible, known as a boat, can hold 10 to 20 pieces of $\sim2$ mm
long and 0.5 mm diameter gold wire. Current flows through the
boat, melting the gold and spewing it upwards toward the
substrate.  A calibrated crystal monitor measures the deposition
rate.  One to two boats are sufficient to deposit 100 to 200 nm of
gold, and this costs \$10 to \$15 per boat. There are typically
only four sets of electrical feedthroughs in the evaporator's
vacuum chamber, and to deposit more gold, one needs to bring the
chamber up to atmosphere, reload the boats with gold, and pump
back down to base pressure---a process that takes about an hour.
The substrate mounting area allows several substrates to be coated
at once. Evaporating less than 1 $\mu$m of gold is reasonable, but
depositing more than 1 $\mu$m becomes too expensive and time
consuming, and the quality of the gold surface begins to diminish.
Moreover, the vacuum chamber eventually becomes hot which may
result in the failure of the crystal monitor or the burning of
photoresist.

\subsection{Photoresist spinning and baking}
\noindent
Photoresist does not always adhere well to the
substrate's surface. Before coating with photoresist, the
substrate should be baked on a hot plate at $\sim150^\circ$C for a
few minutes to remove surface moisture. Hexamethyldisilazane
(HMDS) should be used with sapphire substrates to promote adhesion
(this is unnecessary for AlN). Only a few monolayers of HMDS are
required: after baking, place the sapphire in a dish next to
several drops of HMDS and cover for a few minutes. Note that both
HMDS and photoresist are carcinogenic and should be handled with
care.

Spinning photoresist onto a substrate is a relatively
straightforward process. The substrate, with beads of photoresist
dripped onto its surface, is spun by a vacuum chuck to a few
thousand rpm for several tens of seconds. A faster rotation
results in a thinner film of photoresist. Typically, a film
thickness of a few microns is possible with standard photoresists,
and there exists special resists that are four to twenty microns
thick. These thick resists are often important for making tall
wire structures. The thickness of a photoresist may be increased
beyond its specification by dripping resist onto its surface
during rotation. After spin-coating, the photoresist needs to be
baked on a hot plate to prepare the polymer for UV exposure.  The
exact temperature and bake duration are often crucial to the
success of the fabrication.  We would like to note that it is
possible to layer microwire patterns on top of one another by
fabricating each new wire layer on top of a spin-coated insulator
such as polyimide~\cite{Westervelt01}.

\subsection{UV exposure}
\noindent
The central step in photolithography is the UV exposure
of the photoresist.  An instrument known as a mask aligner allows
one to accurately position the photomask flush to the substrate's
photoresist-coated surface, and a built-in UV lamp exposes the
photoresist for a specified amount of time.  Essential for
photomask and substrate registration is an optical microscope
mounted on the mask aligner.  This enables one to simultaneously
view the wire patterns on the mask and the underlying substrate.
Dust particles or scratches often remain on the substrate even
after a thorough cleaning.  If these defects are sparse, then the
substrate may be translated such that the wires avoid all defects.
Aligning the chip's wire pads along one or more edges of the
substrate further constrains the relative position of the
photomask to the substrate. It should be noted that it is
difficult to properly develop the pads (or other wire features)
less than a millimeter from the edge due to photoresist beading.
Certain fabrication recipes require the photoresist to be baked
and exposed again before developing.

It is good practice to clean the chrome photomasks after every
use. Photoresist can stick to the surface, and if left for days,
will produce hard to remove specs that can block the UV light,
creating unwanted features or breaks in the patterned wires.
Immersing in a dish of acetone and rinsing with IPA and methanol
is sufficient for routine cleaning.  Some chrome masks can
withstand ultrasound cleaning as well as being wiped with a soft,
lint-free cloth, and this seems to be the only way to remove
encrusted grime or particulate.

\subsection{Developing}
\noindent To remove the photoresist regions defined by the UV
exposure, the substrate must be immersed and slightly agitated in
a beaker of developer for a few tens of seconds followed by a
water rinse. The exact developing time depends on the previous
fabrication steps, but it is generally possible, especially with
the transparent sapphire substrates, to see a characteristic
change in opacity of the photoresist as it becomes fully
developed.  For instance, when using a positive process, one first
sees the exposed photoresist turn hazy, revealing the wire
pattern. After a few seconds, the hazy region sloughs off exposing
the bare substrate and leaving darker, patterned regions of
photoresist. If a mistake is made at any point in the
photolithography process, the substrate can be reused by removing
the photoresist in a beaker of acetone and cleaning the substrate
as mentioned above, starting with the ultrasound.

\subsection{Ozone dry stripping}
\noindent
Certain fabrication processes require the substrate
surface to be etched in an ozone dry stripper. This uses UV light,
ozone, and heat to remove thin films of unwanted organic material,
photoresist, or HMDS that may prevent the deposition of thermally
evaporated or electroplated gold.

\subsection{Wire contacts}
\noindent Wire bonding and ultrasonic fluxless soldering are
useful methods for attaching macroscopic wires to the substrate's
contact pads. Wire bonding is the standard method for making
contacts to micro- or nanofabricated devices.  The wire bonder
attaches each end of a thin thread of gold wire to a pad using a
heated, ultrasonically vibrating tip.  The thin wire may be
stretched over several millimeters between the pad on the
substrate and a pad on the substrate support structure.  The pads
on the support structure may then be connected to standard wire
contact pins. Because the wire threads are prone to break and
cannot individually support more than a few hundred mA of current,
it is necessary to make several redundant bonds per pad. This
process can be quite time consuming. As an alternative, ultrasonic
soldering irons are capable of attaching regular wires to sapphire
or AlN using fluxless solder. Attaching wires is nearly as simple
as standard soldering, and the fluxless solder is vacuum
compatible to at least 10$^{-9}$ Torr. Unfortunately, the solder
material forms mounds on the substrate's surface that can limit
optical access.

\subsection{The mirror}
\noindent
Finally, we would like to discuss methods for making the
atom chip's surface mirror-like.  The most straightforward method
involves simply patterning gold on the entire chip's surface
except for thin, $>10\ \mu$m, wide gaps around the actual
wires~\cite{Schmiedmayer00b}. This technique does not add any
additional steps to the fabrication procedure, but it does
increase the likelihood that surface defects will short the wires
through contacts to the large mirrored areas.  The mirror gaps
that define the wires imprint defects onto the reflected mirror
MOT beams, but we have nevertheless been able to trap more than a
million cesium atoms with this less than perfect mirror. Another
technique involves coating the chip's surface with an insulator
and then applying a mirror coating.  For example, several layers
of polymethyl methacrylate (PMMA) can be spun onto the substrate.
Swabbing with acetone removes the PMMA covering the wire pads near
the substrate's edge, and the mirror is created by using a mask to
thermally evaporate gold only onto the PMMA-coated region.
Epoxying a silver mirror (with EPO-TEK 353) to the surface also
forms a good mirror, and it eliminates any corrugations on the
mirror surface caused by the underlying wires~\cite{Jakob01a}.
Unfortunately, the minimum distance between the atoms and the
wires is set by the mirror and epoxy thickness.  An improved
mirror can be made by epoxying a dialectic mirror onto the
surface. Vacuums of $2\times10^{-10}$ Torr, in a chamber baked to
150$^\circ$C, have been achieved despite using this glue and
dielectric coating.

\section{Specific fabrication techniques: wet etching, ion milling, lift-off method, and electroplating}
\noindent
The minimum required wire dimensions vary significantly
depending on the the atom chip's application, and an optimal
fabrication technique should be chosen accordingly.  This section
describes the recipe and relative merit of each fabrication
method.

\subsection{Wet etching and ion milling}
\begin{figure} [htbp]
\centerline{\epsfig{file=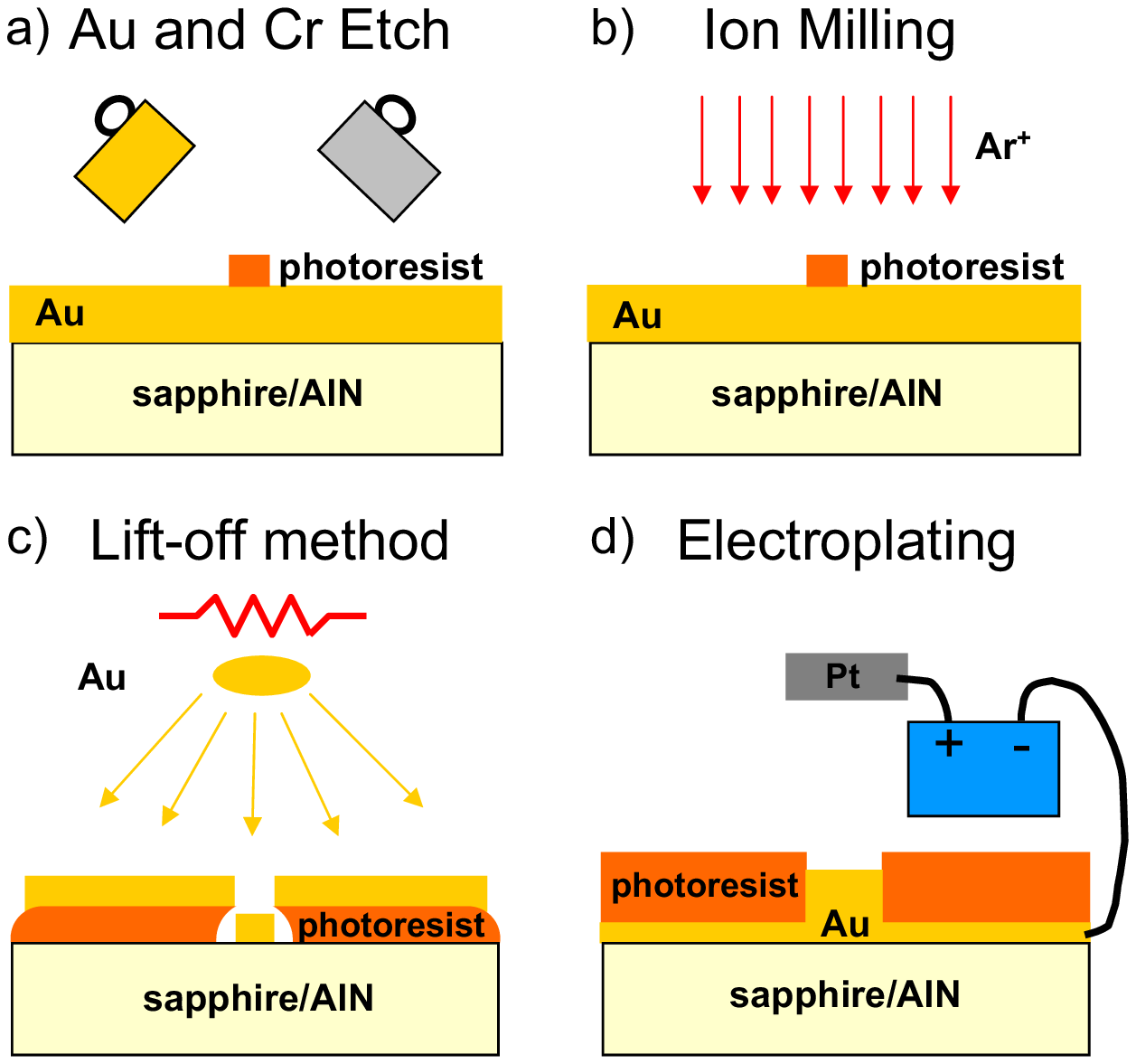, width=11.2cm}}
\vspace*{13pt} \fcaption{\label{fig:fabtech} Fabrication
techniques.  (a) Patterned positive photoresist masks the gold
layer from the gold and chromium wet etch.  (b) The argon ions
mill away the gold not covered by positive photoresist.  (c) Gold
is thermally evaporated into the trenches patterned in the
negative photoresist.  The undercut allows the photoresist and
unwanted gold to separate from the substrate without peeling away
the gold in the trenches. (d) Wires are defined by gaps in the
positive photoresist, and the walls of the photoresist guide the
wires as they are electroplated.  After electroplating, acetone
removes the photoresist and gold and chromium etches remove the
seed layer.}
\end{figure}

\begin{figure} [htbp]
\centerline{\epsfig{file=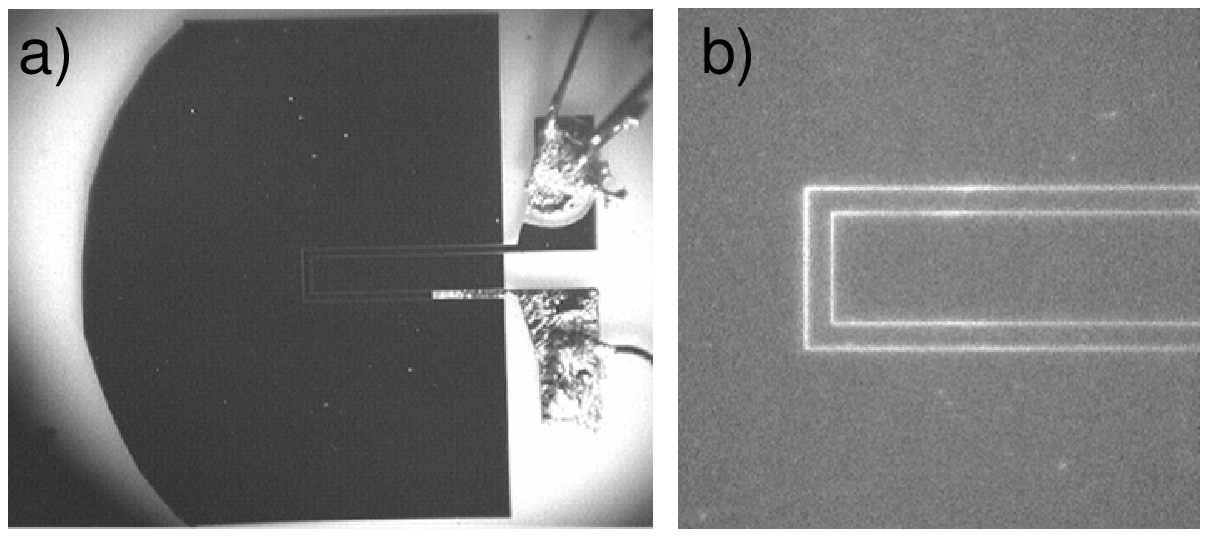, width=11.2cm}}
\vspace*{13pt} \fcaption{\label{fig:WEpics} Gold wire patterned
using the wet etch technique. (a) This atom chip contains a
quadrupole trap in the U configuration. The gold wire, patterned
on sapphire and surrounded by a gold mirror, is 300 $\mu$m wide
and 1 $\mu$m tall. (b) Close-up of the wire region.  The gold
appears darker than the uncovered sapphire substrate.}
\end{figure}

\noindent The simplest chip to fabricate has wire widths no
smaller than 30--40 $\mu$m and wire heights less than 1 $\mu$m.  A
transparency mask should be used for the photolithography.  The
wire height is set by a thermally evaporated gold layer and the
photoresist masks the gold intended for wires from the wet etch
solution (see figure~\ref{fig:fabtech}(a)).
To begin the procedure, the cleaned substrate should be placed in
the ozone dry stripper for five minutes at 65$^\circ$C to ensure
that no organic material will prevent the adhesion of chromium and
gold. The thermal evaporation step follows, with the thickness of
the gold layer determined by chip's current density requirements.
Because the photoresist adheres well to gold, only a 5 min bake at
180$^\circ$C is necessary for adhesion.  Wet etching removes
exposed gold, and the photoresist should be patterned such than it
covers the areas intended for wires, i.e.\ the photoresist should
be a positive image of the wire pattern. A photomask on which the
wires are opaque, used in conjunction with positive photoresist,
will produce a positive image of the wire pattern. We use the
photoresist AZ5214 (Clariant), which can serve as both a negative
and positive photoresist depending on the bake and exposure
procedure.  The positive process recipe is as follows: spin coat
at 5000 rpm for 50 s, bake at 95$^\circ$C for 2 min, expose for 10
to 20 s, and develop in AZ327 MIF (or some similar developer) for
30 s. All of the above times are approximate and will vary
depending on the UV light intensity of the specific mask aligner
and on various environmental conditions. It may be necessary to
try various exposure and bake times to find the optimal recipe.
These exposure times are based on the 16 mW/cm$^2$ UV intensity of
our mask aligner. To remove the gold not covered by photoresist,
submerge the substrate in gold etch solution (Gold Etchant TFA,
Transene Company, Inc., telephone 978-777-7860) for a few tens of
seconds until only the dull gray of the chromium layer remains.
 Finally, remove the chromium layer with chrome etchant (CR-7S,
Cyantek, Co., telephone 510-651-3341). Figures~\ref{fig:WEpics}(a)
and (b) show a substrate patterned in this manner. The wet etch
dissolves the gold isotropically, and the decrease in wire width
is insignificant for wires larger than 10 to 20 $\mu$m. Of course,
transparency masks cannot be used for features smaller than a few
tens of microns.

Ion milling can be useful alternative to wet etching.  Instead of
removing the unwanted gold with an etch solution, argon ions
bombard the surface, removing the gold not covered by photoresist
(see figure~\ref{fig:fabtech}(b)). This method can produce very
narrow features, limited only by photoresist resolution, with
heights determined by the thermally evaporated gold layer.  The
photoresist is also milled, but this is of no consequence as long
as it is thicker than the gold layer. The substrate may become
quite hot during the ion etching, and one needs to be careful that
the substrate does not overheat, causing the photoresist to become
hard and difficult to remove.  We have used ion milling to make
atom chips as well as to etch a common hard drive for use as a
magnetic atom mirror~\cite{Lev03}.

\subsection{The lift-off method}
\noindent The quick and easy wet etch technique is unfortunately
not suitable for wire widths smaller than 10 $\mu$m, and ion
milling machines are not readily available. The lift-off method
should be used for the case in which the wires need not be taller
than 1 $\mu$m but less than 10 $\mu$m wide. In contrast to the wet
etch technique, the photoresist in this method is used as a mask
for the deposition of thermally evaporated gold. Trenches are
created in a negative photoresist using a photomask with opaque
wires, and evaporated gold deposits both into the trenches,
adhering to the substrate, and onto the surface of the photoresist
(see figure~\ref{fig:fabtech}(c)). If done properly, the walls of
the trenches have an overhang---which looks like an undercut when
viewed from above---that prevents the unwanted gold on the
photoresist from connecting to the gold in the trenches. An
acetone bath dissolves the photoresist, allowing the unwanted gold
to lift-off leaving the wire pattern formed from the gold in the
trenches.

After cleaning the substrate, the AZ5214 is spun on the substrate
for 45 s at 5000 rpm.  The maximum height of the thermally
evaporated wires is set by the thickness of the photoresist since
lift-off will not work once the top of the gold connects with the
gold on the overhang.  We have been able to achieve lift-off with
wires 1.5 $\mu$m tall by spinning the photoresist on at 2000 rpm
and thermally evaporating many boats of gold over a period of
three to four hours.  The photoresist should then be baked for 45
s at 100$^\circ$C, UV exposed with the photomask for 10 s, baked
again for 45 s at 123$^\circ$C, UV exposed with no mask for 2.1
min, and developed for 25 to 35 s. Developing is finished when one
can see the wire pattern in the photoresist.  A successful
undercut can be seen in a microscope as a bright outline of the
edges of the trenches. Before thermal evaporation, the substrate
should be placed in the ozone dry stripper at 65$^\circ$C for 5
minutes. To promote lift-off, the acetone bath should be heated on
a hot plate, and the substrate, while inside the beaker, should be
sprayed with an acetone squirt bottle. It is very important that
all of the gold-coated photoresist be pealed away before the
substrate is removed from the acetone. Otherwise, once dried, the
unwanted gold flakes become extremely difficult to separate from
the surface. Difficulty in achieving lift-off may be overcome by
briefly exposing the substrate to ultrasound.  This is risky,
however, since the gold wires might be stripped-off as well.
Figure~\ref{fig:Ioffe}(b) shows an atom chip fabricated with the
lift-off method.

\subsection{Electroplating}
\noindent The above methods rely on thermal evaporation to achieve
the required wire thickness.  This limits the wire heights to
$\sim1\ \mu$m.  Electroplating the wires can increase the wire
height considerably:  for example, we have made 3 $\mu$m wide
wires, 4 $\mu$m tall.  Thick photoresist spun and patterned on a
thin gold seed layer provide a template for the growth of the
wires.  The walls of the photoresist maintain a constant wire
width as the wire height increases (see
figure~\ref{fig:fabtech}(d)). An acetone wash followed by a brief
wet etch removes the photoresist and gold seed layer.
Electroplating is a tricky process that does not always produce
reliable results.  We provide here a general guideline for the
process, and with this process we have typically been able to
achieve a 75\% yield with a wire height accuracy of $\pm0.5\
\mu$m.

Fabrication begins with cleaning and ozone dry stripping the
substrate, followed by the thermal evaporation of a 100 to 150 nm
seed layer of gold along with a 50~\AA~chromium adhesion layer.
For proper wire guiding, the photoresist must always be taller
than the electroplated wires, and a photoresist thicker than the
one used in the aforementioned techniques is necessary. Clariant's
AZ9200 series photoresists are 4 to 24 microns thick, and can
achieve aspect ratios of 5 to 7 with resolutions of $<1\ \mu$m to
3.5 $\mu$m depending on the resist thickness.  After spin coating,
the photoresist should be UV exposed for 60 s (or longer depending
on the photoresist thickness) using a photomask with transparent
wire patterns.  The resist is developed in a 1:4 solution of
AZ400K and water for a minute or more:  the exposed photoresist
will turn hazy before dissolving away.  The gold seed layer also
acts as the cathode in the electroplating process, and some of the
photoresist must be whipped away with acetone---or a blank spot
designed in the photoresist---to serve as a contact for the
cathode lead.  An ozone dry etch is then used to remove any layers
of HMDS, photoresist, or organics that might mask regions of the
gold from the electroplating solution. The time and temperature of
this process is crucial: too long of an exposure at too high of a
temperature will make the photoresist difficult to remove between
closely spaced wires, and too short of an exposure will not remove
enough unwanted masking material.  For example, we found that an
18 s room-temperature ozone dry etch was optimal for removing
unwanted material while also enabling the removal of photoresist
between wires spaced by 3 $\mu$m.

We use a sodium gold sulfite solution (TG-25E, Technic, Inc.
telephone 714-632-0200) for the electroplating.  The solution is
temperature controlled on a hot plate to 60$^\circ$C and agitated
with a magnetic stirrer. The anode is a platinum foil, and the
substrate is connected to the power supply with a standard mini
alligator clip.  This clip can be dipped into the bath to enable
the complete submersion of the substrate.  We usually use a
current of 0.1 to 0.2 mA to electroplate. Higher currents seem to
produce rougher wire surfaces. The solution should remain clear to
slightly yellowish during the process, and something is wrong if
the solution starts to turn brown.  The substrate should be gently
agitated while electroplating to promote even plating and suppress
the formation of $\sim5$ $\mu$m tall towers of gold. Typically, it
takes 10 to 30 minutes to electroplate several microns of gold at
this current setting.

After electroplating, the photoresist should be removed in a
room-temperature acetone bath.  Sometimes it is difficult to
remove the photoresist between wires spaced only several microns
from one another, and in these cases the substrate---while in the
acetone---should be placed in an ultrasound for a few minutes. The
gold should not peel away since it is attached to the entire
substrate surface.  After rinsing the acetone away with IPA and
methanol, the gold seed layer is removed with a $\sim$ 15 s wet
etch. The chromium adhesion layer should also be wet etched away.
Occasionally, the air jet does not remove all of the methanol from
the substrate, and tiny drops of methanol can sometimes dry on
leeward side of the wires.  This dried methanol acts as a mask for
the gold etch, leaving small puddles of the seed layer that can
short adjacent wires.  These puddles can be removed by rinsing
with methanol, blow-drying from a different angle, and briefly wet
etching a second time.  The surface reflectance of the gold is
typically diminished after the wet etch, and a mirror fabricated
with this gold may not be ideal.

\begin{figure} [htbp]
\centerline{\epsfig{file=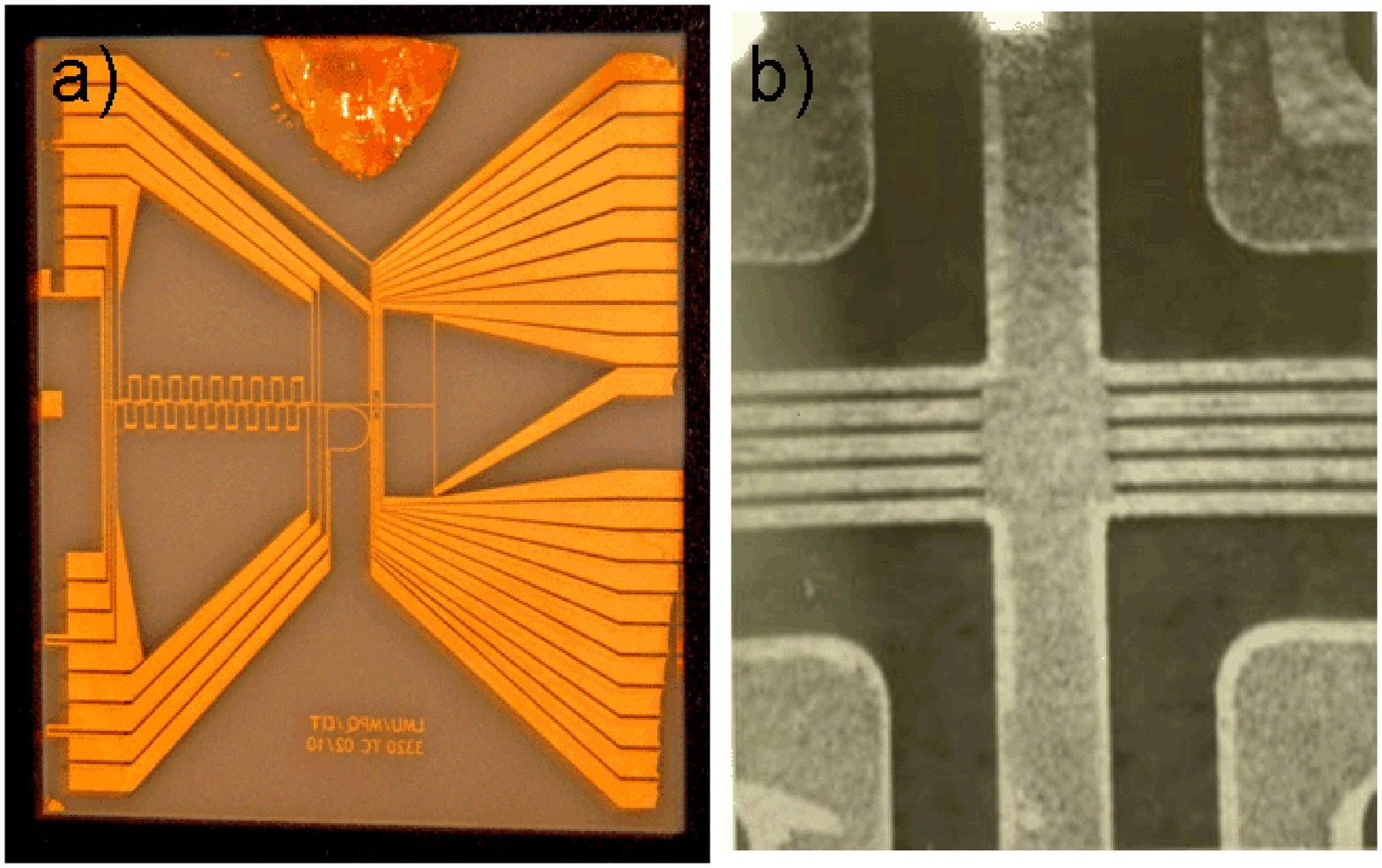, width=10.2cm}}
\vspace*{13pt} \fcaption{\label{fig:BEChip} An atom chip-based BEC
interferometer fabricated by electroplating onto a AlN substrate.
a) The chip will produce a BEC and transport it to the center
region where b) five wires 3 $\mu$m wide, 4 $\mu$m tall, and
spaced by 3 $\mu$m will split the BEC in a double well potential.}
\end{figure}

A surface profilometer, commonly known as an alpha step machine,
is quite useful for quickly measuring the height of the wires.
Inevitably, a few substrates must be spent optimizing the
electroplating process for a specific wire height.
Figures~\ref{fig:BEChip}(a) and (b) show an atom chip-based BEC
interferometer that we fabricated by electroplating on an AlN
substrate~\cite{tilo}. The smallest features are five, 1 mm long
wires that are each 3 $\mu$m wide, 4 $\mu$m tall, and spaced less
than 3 $\mu$m from one another.

\section{Conclusion}
\noindent
The techniques described in this paper provide a basic
starting point for the design and fabrication of these atom chips.
The precise control of atomic position enabled by these chips is
quite crucial to many areas of research.  Moreover, these devices
allow an incredible miniaturization of experimenters involving
cold atoms. From constructing atom optical elements to studies of
BECs and cavity QED, atom chips are proving invaluable to the
fields of atomic physics, quantum optics, and quantum computation.

\nonumsection{Acknowledgements} \noindent  We would like to thank
Michael Roukes and Axel Scherer for use of their clean room
facilities.  Eyal Buks and Darrell Harrington were invaluable in
teaching us the basics of photolithography. We also thank Marko
Loncar for instruction in ion milling, and Jakob Reichel and
Chungsok Lee for their advice regarding electroplating. This work
was performed in the group of Hideo Mabuchi and supported by the
Multidisciplinary University Research Initiative program under
Grant No. DAAD19-00-1-0374.

\nonumsection{References}


\noindent
\begin{thebibliography}{99}

\bibitem{Metcalf99} H. Metcalf and P. van der Straten (1999), {\it Laser Cooling and
Trapping}, Springer-Verlag (New York).

\bibitem{Zoller95} T. Pellizzari, S. A. Gardiner, J. I. Cirac, and
P. Zoller (1995), {\it Decoherence, continuous observation, and
quantum computing: a cavity QED model}, \PRL, {\bf{75}}, 3788.

\bibitem{Calarco} T. Calarco, E. A. Hinds, D. Jaksch, J. Schmiedmayer,
J. I. Cirac, and P. Zoller (2000), {\it Quantum gates with neutral
atoms: Controlling collisional interactions in time-dependent
traps}, \PRA, {\bf{61}}, 022304.

\bibitem{Briegel} R. Raussendorf and H. Briegel (2001), {\it A one-way quantum
computer}, \PRL, {\bf{86}}, 5188.

\bibitem{Libbrecht} J. D. Weinstein and K. G. Libbrecht (1995), {\it Microscopic magnetic traps for neutral atoms}, \PRA,
{\bf{52}}, 4004.

\bibitem{Westervelt98a} M. Drndi\'{c}, K. S. Johnson, J. H.
Thywissen, M. Prentiss, and R. M. Westervelt (1998), {\it
Micro-electromagnets for atom minipulation}, Appl. Phys. Lett.,
{\bf{72}}, 2906.

\bibitem{Schmiedmayer00b} R. Folman, P. Kr\"{u}ger, D. Cassettari, B. Hessmo,
T. Maier, and J. Schmiedmayer (2000), {\it Controlling cold atoms
using nanofabricated surfaces: atom chips}, \PRL, {\bf{84}}, 4749.

\bibitem{Jakob01a} J. Reichel, W. H\"{a}nsel, P. Hommelhoff, and T. W. H\"{a}nsch (2001), {\it Applications of integrated magnetic microtraps},
 Appl. Phys. B, {\bf{72}}, 81.

\bibitem{Folman02} R. Folman, P. Kr\"{u}ger, J. Schmiedmayer, J. Denschlag, and C. Henkel
(2002), {\it Microscopic atom optics: From wires to an atom chip}, Adv. At. Mol. Opt. Phys., {\bf{48}}, 263.

\bibitem{Anderson01} D. M\"{u}ller, E. Cornell, M. Prevedelli, P. Schwindt, Y. Wang, and D. Anderson (2001),
 {\it Magnetic switch for integrated atom optics}, \PRA, {\bf{63}}, 041602.

\bibitem{Hindsrev} E. A. Hinds and I. G. Hughes (1999), {\it Magnetic atom optics: mirrors, guides, traps, and chips for atoms},
\JPD, {\bf{32}}, R119.

\bibitem{Westervelt98b} K. S. Johnson, M. Drndic, J. H. Thywissen, G. Zabow, R. M. Westervelt, and M. Prentiss
(1998), {\it Atomic deflection using an adaptive
microelectromagnet mirror}, \PRL, {\bf{81}}, 1137.

\bibitem{Hannaford02} A. I. Sidorov, R. J. McLean, F. Scharnberg, D. S. Gough, T. J. Davis, B. J. Sexton,
G. I. Opat, and P. Hannaford (2002) {\it Permanent-magnet
microstructures for atom optics}, Acta Physica Polonica B,
{\bf{33}}, 2137.

\bibitem{Lev03} B. Lev, Y. Lassailly, C. Lee, A. Scherer, and H. Mabuchi (2003),
 {\it An atom mirror etched from a hard drive}, quant-ph/0304003.

\bibitem{Levpre} B. Lev and H. Mabuchi (2003), in preparation.

\bibitem{Hinds02} P. Horak, B. Klappauf, A. Haase, R. Folman, J. Schmiedmayer, P. Domokos, and E. A. Hinds
(2002), {\it Towards single-atom detection on a chip},
quant-ph/0210090.

\bibitem{Jakob03} R. Long, T. Steinmetz, P. Hommelhoff, W. H\"{a}nsel, T. W. H\"{a}nsch, and J. Reichel
(2003), {\it Magnetic microchip traps and single-atom detection},
Phil. Trans. R. Soc. Lond. A, {\bf{361}}, 1.

\bibitem{Wineland02a} D. Kielpinski, C. Monroe, and D. J. Wineland
(2002), {\it Architecture for a large-scale ion-trap quantum
computer}, Nature, {\bf{417}}, 709.

\bibitem{Wineland02b} M. A. Rowe, A. Ben-Kish, B. Demarco, D. Leibfried, V. Meyer,
J. Beall, J. Britton, J. Hughes, W. M. Itano, B. Jelenkovi\'{c},
C. Langer, T. Rosenband, and D. J. Wineland (2002), {\it Transport
of quantum states and seperation of ions in a dual RF ion trap},
Quantum Inf. Comput., {\bf{2}}, 257.

\bibitem{Jakob01b} W. H\"{a}nsel, P. Hommelhoff, T. W. H\"{a}nsch, and J. Reichel (2001), {\it Bose-Einstein condensation on a microelectronic chip}
, Nature, {\bf{413}}, 498.

\bibitem{Zimmermann01} H. Ott, J. Fortagh, G. Schlotterbeck, A. Grossmann, and C. Zimmermann (2001),
 {\it Bose-Einstein condensation in a surface microtrap}, \PRL, {\bf{87}}, 230401.

\bibitem{Ketterle03} A. E. Leanhardt, Y. Shin, A. P. Chikkatur, D. Kielpinski, W. Ketterle, and D. E. Pritchard,
(2003), {\it Bose-Einstein condensates near a microfabricated
surface}, \PRL, {\bf{90}}, 100404.

 \bibitem{Schmiedmayer03} R. Folman and J. Schmiedmayer (2003), private
communication.

\bibitem{Schmiedmayer00} D. Cassettari, B. Hessmo, R. Folman, T. Maier, and
J. Schmiedmayer (2000), {\it Beam splitter for guided atoms},
\PRL, {\bf{85}}, 5483.

\bibitem{Mabuchi01} H. Mabuchi, M. Armen, B. Lev, M. Loncar, J. Vu\v{c}kovi\'{c},
H. J. Kimble, J. Preskill, M. Roukes, and A. Scherer (2001), {\it
Quantum networks based on cavity QED}, Quantum Inf. Comput.,
{\bf{1}}, 7.

\bibitem{McKeever03} J. McKeever, J. R. Buck, A. D. Boozer, A. Kuzmich, H. C. N\"{a}gerl, D. M. Stamper-Kurn,
and H. J. Kimble (2003), {\it State-insensitive cooling and
trapping of single atoms in an optical cavity}, \PRL, {\bf{90}},
133602.

\bibitem{Jakob02} J. Reichel (2002), {\it Microchip traps and Bose-Einstein
condensation}, Appl. Phys. B, {\bf{75}}, 469.

\bibitem{Jakob99} J. Reichel, W. H\"{a}nsel, and T. W. H\"{a}nsch (1999),
{\it Atomic micromanipulation with magnetic surface traps}, \PRL,
{\bf{83}}, 3398.

\bibitem{Westervelt01} M. Drndi\'{c}, C. S. Lee, and R. M.
Westervelt (2001), {\it Three-dimensional microelectromagnet traps
for neutral and charged particles}, Phys. Rev. B, {\bf{63}},
085321.

\bibitem{tilo} Wire pattern designed by T. Steinmetz and P.
Hommelhoff at the Max-Plank-Institut f\"{u}r Quantenoptik in
Garching and the Ludwig-Maximilians-Universit\"{a}t in Munich.

\end{thebibliography}
\end{document}